\documentclass[runningheads,a4paper]{llncs}
\pdfoutput=1
\authorrunning{Philipp Jordan and Paula Alexandra Silva}
\usepackage[utf8]{inputenc}
\usepackage{amssymb}
\usepackage{graphicx}
\usepackage[normalem]{ulem}
\useunder{\uline}{\ul}{}
\usepackage[hidelinks]{hyperref}
\hypersetup{
    colorlinks=true,
    linkcolor=blue,
    filecolor=magenta,      
    urlcolor=cyan,
}
\title{Building an Argument for the Use of\\ Science Fiction in HCI Education\thanks{Accepted for publication at \href{http://ihsint.net/index.html}{IHSI 2019}.}}

\titlerunning{Science Fiction in HCI Education}

\author{Philipp Jordan\textsuperscript{1} \and Paula Alexandra Silva\textsuperscript{2}}
\institute{\textsuperscript{1} University of Hawai`i at M{\=a}noa, Honolulu, USA\\
\email{\href{mailto:philippj@hawaii.edu}{philippj@hawaii.edu}} \\
\textsuperscript{2} DigiMedia Research Center, University of Aveiro, Aveiro, Portugal\\
\email{\href{mailto:palexa@gmail.com}{palexa@gmail.com}}}

\begin{document}
\toctitle{IHSINT - arXiv PrePrint}
\tocauthor{Philipp Jordan and Paula Alexandra Silva}
\maketitle

\begin{abstract}
Science fiction literature, comics, cartoons and, in particular, audio-visual materials, such as science fiction movies and shows, can be a valuable addition in Human-computer interaction (HCI) Education. In this paper, we present an overview of research relative to future directions in HCI Education, distinct crossings of science fiction in HCI and Computer Science teaching and the Framework for 21\textsuperscript{st} Century Learning.  Next, we provide examples where science fiction can add to the future of HCI Education. In particular, we argue herein first that science fiction, as tangible and intangible cultural artifact, can serve as a trigger for creativity and innovation and thus, support us in exploring the design space. Second, science fiction, as a means to analyze yet-to-come HCI technologies, can assist us in developing an open-minded and reflective dialogue about technological futures, thus creating a singular base for critical thinking and problem solving. Provided that one is cognizant of its potential and limitations, we reason that science fiction can be a meaningful extension of selected aspects of HCI curricula and research.


\end{abstract}

\section{The Future of HCI Education}
In a 2016 summary article, Churchill, Bowser and Preece \cite{Churchill:2016:FHE:2897145.2888574} outline current priorities in Human-computer interaction (HCI) Education. Among others, the article discusses a  four-year long initiative \cite{CHI_2014_ed_PJ_report} by the Special Interest Group on Computer--Human Interaction (SIGCHI), which assessed the needs and requirements of the current and future HCI curriculum subjects.

Using cross-cultural survey data from a broad, international sample, interview studies, results from discussions at CHI workshops and town-hall conferences, the SIGCHI HCI Education project elicited a total of 114 discrete topics in HCI instruction. In addition, the project created a repository of contemporary HCI courses and curricula \cite{CHI_2014_ed_syllabi}, across a variety of universities and departments.

The survey respondents -- a mix of international HCI professors, practitioners and students  \cite[pages 7-8]{CHI_2014_ed_PJ_report} -- named design research in addition to qualitative evaluation methods, as some of the top priorities in HCI research. The SIGCHI HCI Education project also created an organized and representative overview of relevant HCI classes and syllabi \cite{CHI_2014_ed_syllabi}, including i) Introductory HCI classes, as well as courses on ii) Design, Values \& Ethics or iii) Artificial Intelligence (AI). 

\section{Science Fiction, 21\textsuperscript{st} Century Skills and HCI Education}
\subsubsection{Science Fiction} Science fiction, including Design Fiction\footnote{Design Fiction is an emerging method in Design Research, see for example \cite{Lindley.2016}.}, in its diverse variations and media forms -- written, illustrated, audio-visual or interactive; short story, cartoon or illustrated comic; video clip, show or movie; board or video-game  -- can be a valuable addition in the future of HCI Education. In Michalsky's essay \cite[page 248]{Michalsky.1979} from 1979, the author concludes that `speculative fiction' is:
\begin{quote}
    \textit{``[...] a tool in coping with the onrushing future. Studying speculative fiction offers the student the opportunity to be more creative in his thinking about the future and thus augment the options for possible tomorrow.''}
\end{quote}

The use of science fiction in educational contexts and classrooms has been both, a subject of debate and research in Computer Science and a diversity of other related Science, Technology, Engineering and Mathematics (STEM) fields \cite{Vrasidas.2015}. While the thought of a `science fiction-inspired HCI curriculum and research agenda' \cite{Mubin:2016} is still a pipe dream, noteworthy efforts\footnote{E.g. the ASU Center for Science and the Imagination \cite{ASU} or the `Exchange' \cite{NationalAcademyofSciences.2018}.} have been made to integrate science fiction into HCI-relevant classes across universities in the United States (see Table \ref{hci_classes}); thus acknowledging the educational potential of science fiction. 



\begin{table}[]
\caption{HCI-relevant classes which use aspects of science fiction.}
\label{hci_classes}
\resizebox{\textwidth}{!}{%
\begin{tabular}{|l|l|l|}
\hline
\textbf{Course Alpha} & \textbf{Course Title} & \textbf{Institution} \\ \hline
CS 190 \cite{emory} & Robotics Freshman Seminar & Emory  University \\ \hline
CS 201 \cite{MSU} & AI and Science Fiction & Minnesota State University \\ \hline
CS 463 \cite{Goldsmith:2014:FIC:2600089.2576873} & Introduction to AI and Science Fiction & University of Kentucky \\ \hline
CS 585 \cite{CS_585} & Science Fiction and Computer Ethics & University of Kentucky \\ \hline
STS 1500 \cite{STS1500_2500} & \begin{tabular}[c]{@{}l@{}}Science, Technology and Contemporary Issues:\\  Considering the Future through Fiction \end{tabular} & University of Virginia \\ \hline
STS 2500 \cite{STS1500_2500} & \begin{tabular}[c]{@{}l@{}}Science Fiction and the Future:\\ The Frankenstein Myth in Emerging Biotechnology\end{tabular} & University of Virginia \\ \hline
MAS S64 \cite{MASS64} & Sci Fab: Science Fiction-Inspired Prototyping & Massachusetts Institute of Technology \\ \hline
MAS S65 \cite{Brueckner.2013} & Science Fiction to Science Fabrication & Massachusetts Institute of Technology \\ \hline
\end{tabular}%
}
\end{table}


\subsubsection{21\textsuperscript{st} Century Skills}
The Partnership for 21\textsuperscript{st} Century Skills (P21) \cite{Stanley} is widely recognized as the crucial and visionary framework in teaching and development of knowledge for the next-generation of students and teachers. In a nutshell, the P21 framework \cite{P21-Framework} contains of twelve skills, categorized into three larger domains: i) \textsc{Learning and Innovation}, ii) \textsc{Digital Literacy}, and iii) \textsc{Career and Life Skills}. In particular, the `4C skills' in the \textsc{Learning and Innovation} domain, are considered as \textit{the} profound future learning competencies by the National Education Association and former President Obama \cite[page 5]{NEA_Obama}. 



\subsubsection{Creativity \& Innovation vs. Critical Thinking \& Problem Solving}
In this paper, we argue that two of those four competences -- \textit{Creativity \& Innovation} and \textit{Critical Thinking \&  Problem Solving} -- can be nurtured by resorting to science fiction in  the context of HCI Education.
While both skills, \textit{Creativity \& Innovation} and \textit{Critical Thinking \& Problem Solving}, are viewed independently in the P21 framework, they are naturally interrelated, with the former usually being the preceding cognitive process to the latter. Lin, Tsai, Chien and Chang \cite[pages 198-199]{Lin.2013} provide an example of the co-occurrence of both plus the finding, that science fiction films in the context of:
\begin{quote}
    \textit{``[...] practical educational activities can stimulate students’ imaginations and enhance their ability to design product improvements.''}
\end{quote}


While the first part of above quote highlights how science fiction clearly sparks \textit{Creativity \& Innovation}, the remainder underlines how one needs to engage in a critical reflection process -- \textit{Critical Thinking \& Problem Solving} -- in order to be able to effectively achieve product improvements. Thus, according to the P21 framework, \textit{Creativity \& Innovation} can be seen as \cite[page 1]{P21_Creativity_and_Innovation}:
\begin{quote}
    \textit{``the ability to produce and implement new, useful ideas [...].''} 
\end{quote}
Due to it's strong -- in the case of science fiction movie and shows, audio-visual -- context, embedded in a rich narrative, science fiction has the potential to allow  students and educators to explore the full bandwidth of the design space; from positive (utopian) to negative (dystopian) future visions.

Despite of frequently representing `unrealistic' and `technologically implausible' futures, science fiction materials can serve as creativity triggers and be  pivotal to uncover new design possibilities. For example, a fictional robot from the Disney movie Big Hero 6 \cite{Big_Hero_6}, called BayMax, has inspired researchers to create (and evaluate) a real-world care robot called `Puffy' -- an innovative companion for children with neurodevelopmental disorder \cite{gelsomini2017puffy}. 


\textit{Critical thinking \& Problem Solving} can concisely be defined as \cite[page 1]{P21_Critical_Thinking}: 
\begin{quote}
    \textit{{``strategies we use to think in organized ways to analyze and solve problems [...]'.''}}
\end{quote}\textit Again, due to its rich context and narrative, but also as a cultural artifact, science fiction can be useful in outlining ethical concerns, and to spark dialogue, analysis and reflection about yet-to-come interfaces, interactions, devices and technological outcomes. Uncovering potential moral and societal implications can be particularly interesting when the discussion is led by questions on the how, when, whys and why nots a given Human-Technology interaction is acceptable, plausible, etc. By coupling science fiction examples with prompt questions which lead to reflection, we can make our knowledge and concerns explicit and thus,  adjust to future design endeavors. A similar strategy has been previously applied in educational and creativity contexts \cite{Silva:2010:BMC:1854969.1854992}.

Science fiction has been used in teaching computer ethics \cite{Burton:2018:TCE:3241891.3154485} and computer security \cite{Kohno.2011} classes, therefore encouraging alternative viewpoints and extending traditional technical foci in HCI Education. Rogers \cite[page 679]{Rogers.2010} refers to science fiction movies and shows as \textit{``culturally current media''}, which can not only illustrate good and bad user interface design, but also support the development of  \textit{``design strategies, application and evaluation''} for innovation in HCI curricula.
\section{Discussion and Concluding Remarks}
Science fiction in HCI research has been discussed prior \cite{Jordan2017,Jordan_2018}. The value of fictional visions of the future was introduced as early as 1992 in a CHI panel \cite{Marcus.1992} -- about 25 years ago. Science fiction can be used to imagine the tomorrow \cite{Marcus.2012}, to inspire HCI research \cite{Russell:2018:WLS:3190768.3178552} and user interface design \cite{Shedroff.2012,Figueiredo.2015}. Furthermore, science fiction can showcase future technologies \cite{Schmitz.2008} well ahead of time and reportedly  stimulated research and development in medical device-  \cite{XPRIZEFoundation.2018} or robot-design \cite{puffy_youtube}.

However, science fiction is by no means a `cure-all' for HCI Education. For example, Lin et al. \cite{Lin.2013} and Barnett et al. \cite{Barnett.2006} found that science fiction movies do have an impact on the understanding and perception of students on scientific mechanisms and concepts -- positively or negatively. In addition, Myers and Abd-El-Khalick \cite{Myers.2016} provide a classroom example, where the assumptions in a science fiction film eventually lead students towards detrimental learning outcomes.

In this paper, we endeavored to highlight intersections of science fiction and HCI Education, in view of two crucial skills of the P21 framework. Although, past research concerning science fiction materials in educational contexts has shown mixed results, we reason that science fiction can be of high-value in classroom settings and the future of HCI Education; when utilized with fore-thought and due care. Through such a `conscientious' integration of science fiction in HCI activities and syllabi, we reason that the benefits will outweigh the drawbacks and possibly pave the way toward a science fiction-inspired HCI curriculum.

\bibliographystyle{splncs03}
\bibliography{splncs03}

\end{document}